\documentclass[12pt]{article}
\usepackage{graphicx}
\def \babar{B{\sc a}B{\sc ar}}
\def\beq{\begin{equation}}
\def\eeq{\end{equation}}
\def\bea{\begin{eqnarray}}
\def\eea{\end{eqnarray}}
\def\bwt{\begin{widetext}}
\def\ewt{\end{widetext}}
\def\nn{\nonumber}
\def\roughly#1{\mathrel{\raise.3ex\hbox
{$#1$\kern-.75em\lower1ex\hbox{$\sim$}}}}
\def\lsim{\roughly<}

\def\kbar{{\bar K}^0}
\def\bd{B^0}
\def\bs{B_s^0}

\def\piz{\pi^0}
\def\pip{\pi^+}
\def\pim{\pi^-}
\def\pewc{P_{EW}^C}
\def\pew{P_{EW}}

\def\btod{{\bar b} \to {\bar d}}
\def\btos{{\bar b} \to {\bar s}}
\def\order{\lower 1.8ex \hbox{\LARGE\~{}}}

\def\btokpipi{B \to K \pi \pi}
\def\btokkk{B \to KK{\bar K}}

\def \ll{\left|}
\def \rr{\right|}
\def \<{\left<}
\def \>{\right>}
\def \[{\left[}
\def \]{\right]}
\def \({\left(}
\def \){\right)}
\def \lb{\left\{}
\def \rb{\right\}}
\def \l.{\left.}
\def \r.{\right.}
\def \nl{\nonumber\\}
\def \s{\sqrt{2}}
\def \st{\sqrt{3}}
\def \sf{\sqrt{5}}
\def \sx{\sqrt{6}}
\def \hf{\frac{1}{2}}
\def \thf{\frac{3}{2}}

\def \oth{\frac{1}{3}}
\def \tth{\frac{2}{3}}

\def \ok{\overline{K}^0}
\def \FS{{(fs)}}

\begin{document}

\begin{flushright}
UdeM-GPP-TH-14-230 \\
TECHNION-PH-14-3 \\
EFI 14-3 \\
\end{flushright}

\begin{center}
\bigskip
{\Large \bf \boldmath Charmless $B\to PPP$ Decays: \\ the Fully-Symmetric Final State} \\
\bigskip
\bigskip
{\large
Bhubanjyoti Bhattacharya $^{a,}$\footnote{bhujyo@lps.umontreal.ca},
Michael Gronau $^{b,}$\footnote{gronau@physics.technion.ac.il},
Maxime Imbeault $^{c,}$\footnote{mimbeault@cegep-st-laurent.qc.ca},
David London $^{a,}$\footnote{london@lps.umontreal.ca},
and Jonathan L. Rosner $^{d,}$\footnote{rosner@hep.uchicago.edu}
}
\end{center}

\begin{flushleft}
~~~~~~~~~~~$a$: {\it Physique des Particules, Universit\'e
de Montr\'eal,}\\
~~~~~~~~~~~~~~~{\it C.P. 6128, succ. centre-ville, Montr\'eal, QC,
Canada H3C 3J7}\\
~~~~~~~~~~~$b$: {\it Physics Department, Technion -- Israel Institute of Technology,}\\
~~~~~~~~~~~~~~~{\it Haifa 3200, Israel}\\
~~~~~~~~~~~$c$: {\it D\'epartement de physique, C\'egep de Saint-Laurent,}\\
~~~~~~~~~~~~~~~{\it 625, avenue Sainte-Croix, Montr\'eal, QC, Canada H4L 3X7 } \\
~~~~~~~~~~~$d$: {\it Enrico Fermi Institute and Department of Physics,}\\
~~~~~~~~~~~~~~~{\it University of Chicago, 5620 S. Ellis Avenue, Chicago, IL 60637 }
\end{flushleft}

\begin{center}
\bigskip (\today)
\vskip0.5cm {\Large Abstract\\} \vskip3truemm
\parbox[t]{\textwidth}{ In charmless $B\to PPP$ decays, where $P$ is a
  pseudoscalar meson, there are six possibilities for the symmetry of
  the final state. In this paper, for $P=\pi,K$, we examine the
  properties of the fully-symmetric final state. We present
  expressions for all 32 $B\to PPP$ decay amplitudes as a function of
  both SU(3) reduced matrix elements and diagrams, demonstrating the
  equivalence of diagrams and SU(3). We also give 25 relations among
  the amplitudes in the SU(3) limit, as well as those that appear when
  the diagrams $E$/$A$/$PA$ are neglected.  In the SU(3) limit, one
  has the equalities $\sqrt{2} {\cal A}(B^+\to K^+\pi^+\pi^-)_{\rm FS}
  = {\cal A}(B^+\to K^+K^+K^-)_{\rm FS}$ and $\sqrt{2} {\cal A}(B^+\to
  \pi^+ K^+K^-)_{\rm FS} = {\cal A}(B^+\to \pi^+\pi^+\pi^-)_{\rm FS}$,
  where FS denotes the fully-symmetric final state.  These provide
  good tests of the standard model that can be carried out now by the
  LHCb Collaboration.  }

\end{center}


\thispagestyle{empty}
\newpage
\setcounter{page}{1}
\baselineskip=14pt

\section{INTRODUCTION}

Over the past several years, a certain amount of attention, both
theoretical and experimental, has focused on charmless, three-body
$B\to PPP$ decays ($P$ is a pseudoscalar meson). First, in
Ref.~\cite{3body1} the diagrammatic method was proposed to describe
$B\to PPP$ decays. Using this, it was shown that clean information
about weak Cabibbo-Kobayashi-Maskawa (CKM) phases can be extracted
from three-body decays \cite{3body2,3body3}. This idea was applied in
Ref.~\cite{BIL} to obtain the weak phase $\gamma$ from the
\babar\ measurements of the Dalitz plots for some $\btokpipi$ and
$\btokkk$ decays. Second, the LHCb collaboration recently reported
nonzero measurements of CP asymmetries for the $\Delta S=1$ decays
$B^+ \to K^+\pi^+\pi^-$ and $B^+ \to K^+K^+K^-$ \cite{LHCbDS1}, as
well as for the $\Delta S=0$ decays $B^+ \to \pi^+\pi^+\pi^-$ and
$B^+\to \pi^+K^+K^-$ \cite{Aaij:2013bla}. Furthermore, considerably
larger CP asymmetries were measured in these decays for localized
regions of phase space. That is, the CP-asymmetry measurements are
momentum dependent.  Theoretical analyses of these results can be
found in Refs.~\cite{Bhattacharya:2013cvn,He,MG,Cheng}.

For charmless $B\to PPP$ decays, under flavor SU(3) the three
final-state particles are treated as identical, so that the six
permutations of these particles must be considered.  There are thus
six possibilities for the final state: a totally symmetric state, a
totally antisymmetric state, or one of four mixed states. The
fully-symmetric state is particularly intriguing for the following
reason. The state that is fully symmetric under permutations of the
particle indices is, by definition, also fully symmetric under
permutations of the particle momenta. As such, this state does not
receive contributions from spin-1 resonances because the decay
products of such resonances are necessarily in a state antisymmetric
in the particle momenta. This means that all effects generated by
spin-1 resonances, such as direct CP asymmetries, SU(3) breaking,
etc., are absent in the fully-symmetric $PPP$ state.

In this paper we examine in detail the properties of the
fully-symmetric final state in $B\to PPP$. For simplicity, we assume
that the final-state particles are all $\pi$'s or $K$'s. We first
establish which SU(3) reduced matrix elements contribute to the
fully-symmetric state, and write all $B\to PPP$ decay amplitudes as a
function of these matrix elements. It turns out that there are seven
independent combinations of matrix elements. However, there are a
total of 32 decays -- 16 $\btos$ and 16 $\btod$ transitions. There are
therefore 25 relations among the amplitudes in the SU(3) limit. A
number of these are subject to experimental tests. If the relations
are found not to hold, this would indicate SU(3) breaking, or possibly
even new physics.

We also write all $B\to PPP$ decay amplitudes as a function of
diagrams. A comparison of the two expressions for the amplitudes
allows us to write the SU(3) matrix elements as a function of
diagrams, which establishes the equivalence of diagrams and SU(3), as
was also done for $B\to PP$ decays \cite{GHLR1}. As in two-body
decays, three of the diagrams -- $E$, $A$ and $PA$ -- involve
interactions of the spectator quark, and are expected to be quite a
bit smaller than the other diagrams. In the limit in which $E$, $A$
and $PA$ are neglected, the number of combinations of matrix elements
is reduced from seven to five. This implies additional relations among
the amplitudes, and several of these can also be tested
experimentally.

All experimental tests require that the fully-symmetric final state be
probed. For any $B\to PPP$ decay, this can be done using an isobar
analysis of the Dalitz plot. This was discussed in Ref.~\cite{3body1},
and we review the method later in the paper. But the point is that,
for any three-body decay for which a Dalitz plot has been measured,
one can extract the fully-symmetric amplitude.

One experimental test that is quite compelling, and which can be
performed now, is related to the above LHCb measurements of CP
asymmetries. In the SU(3) limit, one has the equalities $\sqrt{2}
{\cal A}(B^+\to K^+\pi^+\pi^-)_{\rm FS} = {\cal A}(B^+\to
K^+K^+K^-)_{\rm FS}$ and $\sqrt{2} {\cal A}(B^+\to \pi^+K^+K^-)_{\rm
  FS} = {\cal A}(B^+\to \pi^+\pi^+\pi^-)_{\rm FS}$ (FS stands for
fully symmetric). This says that the amplitudes for the
fully-symmetric state of the two decays in each equality are predicted
by the standard model (SM) to be equal. Furthermore, these equalities
are momentum dependent, so that one should find the same values for
the amplitudes at all points in the Dalitz plot.  Now, LHCb has
already measured the Dalitz plots for these decays.  By performing an
isobar analysis, the fully-symmetric amplitudes can be
constructed. These relations can then be examined, providing a test of
the SM.

In Sec.~II, we express the fully-symmetric amplitudes for all $B\to
PPP$ decays in terms of SU(3) reduced matrix elements. We present the
relations among these amplitudes in Sec.~III. In Sec.~IV, we express
the fully-symmetric $B\to PPP$ decay amplitudes in terms of diagrams,
and demonstrate the equivalence of SU(3) and diagrams. We discuss the
experimental tests of the amplitude relations in Sec.~V, with an
emphasis on the relations involving $B^+ \to K^+\pi^+\pi^-$ and $B^+
\to K^+K^+K^-$, and $B^+ \to \pi^+\pi^+\pi^-$ and $B^+ \to
\pi^+K^+K^-$. We conclude in Sec.~VI.

\section{AMPLITUDES \& REDUCED MATRIX ELEMENTS}

In this section we perform SU(3) Wigner-Eckart decompositions of the
$B\to PPP$ decay amplitudes. Each element of SU(3) can be represented
by $\ll{\bf r} Y I I_3\>$, where ${\bf r}$ is the irreducible
representation (irrep), $Y$ is the hypercharge, and $I$ and $I_3$
stand for isospin and its third component, respectively. Note that, in
general, Lie algebras are not associative, so that the order of
multiplication is important. Here we take products from left to right.

In order to construct products of SU(3) states we use the SU(3)
isoscalar factors from Refs.~\cite{deSwart:1963,Kaeding}, along with
SU(2) Clebsch-Gordan coefficients (CG's). We have checked that the
results match with those SU(3) CG's that are listed in
Ref.~\cite{McNamee:1964}.

There are 16 $\btos$ and 16 $\btod$ charmless three-body $B\to PPP$
decays, where $P = \pi$ or $K$. Under flavor SU(3), all three final%
-state particles belong to the same multiplet (octet of SU(3)), and
hence they can be treated as identical, so that the six possible
permutations of these particles must be considered. Here we focus
on the fully-symmetric final state, which has dimension 120. This
can be decomposed into irreps of SU(3) as follows:
\bea
({\bf 8}\times{\bf 8}\times{\bf 8})_{\rm FS} = {\bf 64} + {\bf 27}_{\rm FS} + {\bf 10}_{\rm FS} + {\bf 10^*}_{\rm FS} + {\bf 8}_{\rm FS} + {\bf 1} ~,
\eea
where
\bea
{\bf 27}_{\rm FS} &=& \sqrt{\frac{8}{15}}{\bf 27}_{{\bf 8}\times{\bf 27}} + \sqrt{\frac{7}{15}}{\bf 27}_{{\bf 8}\times{\bf 8}} ~, \nl
{\bf 10}_{\rm FS} &=& \sqrt{\frac{2}{5}}{\bf 10}_{{\bf 8}\times{\bf 27}} + \sqrt{\frac{3}{5}}{\bf 10}_{{\bf 8}\times{\bf 8}} ~, \nl
{\bf 10^*}_{\rm FS} &=& \sqrt{\frac{2}{5}}{\bf 10^*}_{{\bf 8}\times{\bf 27}} - \sqrt{\frac{3}{5}}{\bf 10^*}_{{\bf 8}\times{\bf 8}} ~, \nl
{\bf 8}_{\rm FS} &=& \frac{3}{2\sf}{\bf 8}_{{\bf 8}\times{\bf 27}} - \sqrt{\frac{2}{15}}{\bf 8}_{{\bf 8}\times{\bf 8}} + \sqrt{\frac{5}{12}}{\bf 8}_{{\bf 8}\times{\bf 1}} ~.
\eea
In the first equation above, note that ${\bf 27}_{{\bf 8}\times{\bf
    27}}$ is symmetric under the interchange of ${\bf 8}$ and ${\bf
  27}$. This distinguishes this irrep from ${\bf 27'}$ which also
appears in the Wigner-Eckart decomposition of ${\bf 8}\times{\bf 27}$,
but is antisymmetric under the interchange of ${\bf 8}$ and ${\bf
  27}$. Similarly, in the fourth equation, ${\bf 8}_{{\bf 8}\times{\bf
    8}}$ is symmetric under the interchange of two ${\bf 8}$'s,
distinguishing it from the other ${\bf 8}$ in the product ${\bf
  8}\times{\bf 8}$ that is antisymmetric.

\subsection{SU(3) assignments of pseudoscalar mesons}

We begin by representing the light-quark states ($u$, $d$ and $s$) in
flavor SU(3). The three quarks transform as the fundamental (${\bf 3}$)
of SU(3). The antiquarks transform as the ${\bf 3^*}$ of SU(3). We have
assigned the following representations to the quarks and antiquarks:
\bea
\ll u\> ~=~ \ll{\bf 3}\oth\hf\hf\>~,&~~~~~~&-\ll{\bar u}\> ~=~ \ll{\bf 3^*}{-}\oth\hf{-}\hf\>~, \nl
\ll d\> ~=~ \ll{\bf 3}\oth\hf{-}\hf\>~,&~~~~~~&\ll{\bar d}\> ~=~ \ll{\bf 3^*}{-}\oth\hf\hf\>~, \nl
\ll s\> ~=~ \ll{\bf 3}{-}\tth00\>~,&~~~~~~&\ll{\bar s}\> ~=~ \ll{\bf 3^*}\tth00\>~.
\eea
Using this convention we find that the $\pi$'s and $K$'s can be
represented as follows:
\bea
\ll\pi^+\> ~=~ \ll u\>\ll{\bar d}\>~=~\ll{\bf 8}011\>~,~~~~~\ll\pi^-\> ~=~ -\ll d\>\ll{\bar u}\>~=~\ll{\bf 8}01{-}1\>~,~~~~~~~ \nl
\ll\pi^0\> ~=~ \frac{\ll d\>\ll{\bar d}\> - \ll u\>\ll{\bar u}\>}{\s}~=~\ll{\bf 8}010\>~,~~~~~~~~~~~~~~~~~~~~~~~~~~\nl
\ll K^+\> ~=~ \ll u\>\ll{\bar s}\>~=~\ll{\bf 8}1\hf\hf\>~,~~~~~\ll K^0\> ~=~ \ll d\>\ll{\bar s}\>~=~\ll{\bf 8}1\hf{-}\hf\>~,~~~~~~~ \nl
\ll\ok\> ~=~ \ll s\>\ll{\bar d}\>~=~\ll{\bf 8}{-}1\hf\hf\>~,~~~~~\ll K^-\> ~=~ -\ll s\>\ll{\bar u}\>~=~\ll{\bf 8}{-}1\hf{-}\hf\>~.
\eea

\subsection{\boldmath Fully-symmetric three-body final states}

The first step is to construct normalized three-body states within
flavor SU(3) representing fully-symmetric $P_1P_2P_3$ final states.
There are three cases. In the first, all three particles in the final
state are distinct from one another (e.g., $\pi^0\pi^+\pi^-$).  We
begin by constructing states that are symmetrized over the first two
particles. We then add all three combinations symmetrized in this way
to obtain the fully-symmetric state. In what follows the state is
symmetrized over particles that are included within parentheses.
\bea
\label{eq:sym}
\ll (P_1 P_2) P_3 \> &=& \frac{1}{\s}\[\ll P_1\>\ll P_2\>\ll P_3\> + \ll P_2\>\ll P_1\>\ll P_3\>\] ~,\nl
\ll (P_1 P_2 P_3) \>_{\rm FS} &=& \frac{1}{\st}\[(\ll P_1\>\ll P_2\>)\ll P_3\> + (\ll P_2\>\ll P_3\>)\ll P_1\> + (\ll P_3\>\ll P_1\>)\ll P_2\>\] ~.
\eea
In the second case, two of the three particles are identical but
distinct from the third (e.g., $\pi^0\pi^0\pi^+$). The three-particle
state in which the first two particles are identical is automatically
symmetric in the first two particles. Note that this state cannot be
obtained from the first of Eq.\ (\ref{eq:sym}) above by simply setting
$P_1 = P_2$. In writing Eq.\ (\ref{eq:sym}) above we have used the
fact that the states $\ll P_1 \>$ and $\ll P_2 \>$ are orthogonal to
one another, which is no longer true if they represent the same
particle. Keeping this in mind, we construct the fully-symmetric
three-particle state by adding only distinct combinations symmetrized
over the first two particles.
\bea
\ll (P P) P_3 \> &=& \ll P\>\ll P\>\ll P_3\> ~,\nl
\ll (P P_3) P \> &=& \frac{1}{\s}\[\ll P\>\ll P_3\>\ll P\> + \ll P_3\>\ll P\>\ll P\>\] ~,\nl
\ll (P P P_3) \>_{\rm FS} &=& \frac{1}{\st}(\ll P\>\ll P\>)\ll P_3\> + \sqrt{\frac{2}{3}}(\ll P\>\ll P_3\>)\ll P\> ~.
\eea
The final case is where all three particles are identical (e.g.,
$\pi^0\pi^0\pi^0$). In this case the state obtained by multiplying
three single-particle states is automatically symmetric over all three
particles.  Once again note here that, in order to construct a state
that is appropriately normalized, it is not possible to simply set
$P_1 = P_2 = P_3$ in the cases discussed above.
\bea
\ll (P P P) \>_{\rm FS} &=& \ll P\>\ll P\>\ll P\> ~.
\eea

\subsection{\boldmath Three-body $\btos$ and $\btod$ transitions using flavor SU(3)}

The Hamiltonian for three-body $B$ decays follows from the underlying
quark-level transitions $\btos q{\bar q}$ and $\btod q{\bar q}$, where
$q$ is an up-type quark ($u, c, t$). However, the unitarity of the CKM
matrix, given as
\bea
\sum\limits_{q = u, c, t}V^*_{qb}V^{}_{qs} = 0 ~~,~~~~
\sum\limits_{q = u, c, t}V^*_{qb}V^{}_{qd} = 0~,
\label{unitarity}
\eea
allows us to trade one of the up-type quarks for the other two.  Here
we choose to replace the $t$-quark operators and retain only the
$c$-quark and $u$-quark operators. Thus the weak-interaction
Hamiltonian is composed of four types of operators, namely $\btos
c{\bar c}$, $\btod c{\bar c}$, $\btos u {\bar u}$, and $\btod u{\bar
  u}$. The flavor-SU(3) representations of these operators are
dictated by the final-state light quarks, since the heavy $c$, $b$ and
$t$ quarks are flavor-SU(3) singlets. The transition operators are
given as follows:
\bea
{\cal O}_{\btos c{\bar c}} &=& V^{*}_{cb}V^{}_{cs}B^{(\bf 3^*)}_{\{\tth, 0, 0\}} ~, ~~~~~
{\cal O}_{\btod c{\bar c}} ~=~ V^{*}_{cb}V^{}_{cd}B^{(\bf 3^*)}_{\{-\oth, \hf, \hf\}}  ~, \nl
{\cal O}_{\btos u{\bar u}} &=& V^{*}_{ub}V^{}_{us}\lb A^{(\bf 3^*)}_{\{\tth, 0, 0\}} + R^{(\bf 6)}_{\{\tth, 1, 0\}} + \sx P^{(\bf 15^*)}_{\{\tth, 1, 0\}} + \st P^{(\bf 15^*)}_{\{\tth, 0, 0\}}\rb ~,\nl
{\cal O}_{\btod u{\bar u}} &=& V^{*}_{ub}V^{}_{ud}\lb A^{(\bf 3^*)}_{\{-\oth, \hf, \hf\}} - R^{(\bf 6)}_{\{-\oth, \hf, \hf\}} + \sqrt{8} P^{(\bf 15^*)}_{\{-\oth, \thf, \hf\}} + P^{(\bf 15^*)}_{\{-\oth, \hf, \hf\}}\rb ~,
\eea
where we have used the notation $O^{(\bf r)}_{\{Y, I, I_3\}}$
to represent each flavor-SU(3) operator ($O = \{A, B, R, P\}$).
We have taken the names and relative signs between these operators
from Ref.~\cite{Zeppenfeld}. The weak-interaction Hamiltonian
that governs charmless $B$ decays is then simply a sum of these
four operators:
\bea
{\cal H} &=& {\cal O}_{\btos c{\bar c}} + {\cal O}_{\btod c{\bar c}} + {\cal O}_{\btos u{\bar u}} + {\cal O}_{\btod u{\bar u}}  ~.
\eea
The above Hamiltonian governs the decay of the flavor-SU(3) triplet
of $B$-mesons [$B^{\bf 3} = (B^+_u, B^0_d, B^0_s$)], whose components
have the same flavor-SU(3) representations as their corresponding
light quarks. The fully-symmetric three-body decay amplitude for the
process $B \to P_1 P_2 P_3$ can now be constructed easily as follows:
\bea
{\cal A}_{\rm FS}(p_1, p_2, p_3) &=& _{\rm FS}\<(P_1P_2P_3)\ll\rr{\cal H}\ll\rr B^{\bf 3}\>~,
\eea
where $p_i$ represents the momentum of the final-state particle $P_i$.

\section{AMPLITUDE RELATIONS}

The 32 charmless three-body $B$ decay amplitudes (16 $\btos$ and 16
$\btod$) can all be written in terms of nine matrix elements (here we
have suppressed the $Y, I, I_3$ indices of the operators):
\bea
B^{\FS}_1 &\equiv& _{\rm FS}\<{\bf 1}\ll\ll B^{(\bf 3^*)}\rr\rr{\bf 3}\> ~, \nl
B^{\FS}   &\equiv& _{\rm FS}\<{\bf 8}\ll\ll B^{(\bf 3^*)}\rr\rr{\bf 3}\> ~, \nl
A^{\FS}_1 &\equiv& _{\rm FS}\<{\bf 1}\ll\ll A^{(\bf 3^*)}\rr\rr{\bf 3}\> ~, \nl
A^{\FS}   &\equiv& _{\rm FS}\<{\bf 8}\ll\ll A^{(\bf 3^*)}\rr\rr{\bf 3}\> ~, \nl
R_8^{\FS} &\equiv& _{\rm FS}\<{\bf 8}\ll\ll R^{(\bf 6)}\rr\rr{\bf 3}\> ~, \nl
R_{10}^{\FS} &\equiv& _{\rm FS}\<{\bf 10}\ll\ll R^{(\bf 6)}\rr\rr{\bf 3}\> ~, \nl
P_{10^*}^{\FS} &\equiv& _{\rm FS}\<{\bf 10^*}\ll\ll P^{(\bf 15^*)}\rr\rr{\bf 3}\> ~, \nl
P_8^{\FS} &\equiv& _{\rm FS}\<{\bf 8}\ll\ll P^{(\bf 15^*)}\rr\rr{\bf 3}\> ~, \nl
P_{27}^{\FS} &\equiv& _{\rm FS}\<{\bf 27}\ll\ll P^{(\bf 15^*)}\rr\rr{\bf 3}\> ~.
\eea
The decomposition of all 32 amplitudes in terms of these matrix
elements is given in Tables \ref{tab:1} and \ref{tab:2}.

\begin{table}
\renewcommand*{\arraystretch}{2}
\caption{Amplitudes for $\Delta S = 1$ $B$-meson decays to
  fully-symmetric $PPP$ states as a function of the SU(3) matrix
  elements.
\label{tab:1}}
\begin{center}
\begin{tabular}{|c|c c|c c c c c c c|} \hline \hline
Decay & \multicolumn{2}{c|}{$V^*_{cb}V^{}_{cs}$} & \multicolumn{7}{c|}{$V^*_{ub}V^{}_{us}$} \\ \cline{2-10}
Amplitude & $B^{\FS}_1$ & $B^{\FS}$ & $A^{\FS}_1$ & $A^{\FS}$ & $R^{\FS}_8$ & $R^{\FS}_{10}$ & $P^{\FS}_8$ & $P^{\FS}_{10^*}$ & $P^{\FS}_{27}$ \\ \hline \hline
$A(B^+ \to K^+\pi^+\pi^-)$ & 0 & $\frac{1}{\sqrt{5}}$ & 0 & $\frac{1}{\sqrt{5}}$ & $\frac{1}{\sqrt{15}}$ & $-\frac{1}{3\st}$ & $-\frac{3}{5}$ & 0 & $ - \frac{3\s}{5\sqrt{7}}$ \\
$\s A(B^+ \to K^+\pi^0\pi^0)$ & 0 & $-\frac{1}{\sqrt{5}}$ & 0 & $-\frac{1}{\sqrt{5}}$ & $-\frac{1}{\sqrt{15}}$ & $-\frac{2}{3\st}$ & $\frac{3}{5}$ & 0 & $\frac{18\s}{5\sqrt{7}}$ \\
$\s A(B^+ \to K^0\pi^+\pi^0)$ & 0 & 0 & 0 & 0 & 0 & $\frac{1}{\st}$ & 0 & 0 & $-\frac{3\s}{\sqrt{7}}$ \\
$A(\bd \to K^0\pi^+\pi^-)$ & 0 & $\frac{1}{\sqrt{5}}$ & 0 & $\frac{1}{\sqrt{5}}$ & $-\frac{1}{\sqrt{15}}$ & $\frac{1}{3\st}$ & $\frac{1}{5}$ & $\frac{2\s}{3}$ & $-\frac{9\s}{5\sqrt{7}}$ \\
$\s A(\bd \to K^0\pi^0\pi^0)$ & 0 & $-\frac{1}{\sqrt{5}}$ & 0 & $-\frac{1}{\sqrt{5}}$ & $\frac{1}{\sqrt{15}}$ & $\frac{2}{3\st}$ & $-\frac{1}{5}$ & $-\frac{2\s}{3}$ & $-\frac{6\s}{5\sqrt{7}}$ \\
$\s A(\bd \to K^+\pi^0\pi^-)$ & 0 & 0 & 0 & 0 & 0 & $-\frac{1}{\st}$ & 0 & 0 & $\frac{3\s}{\sqrt{7}}$ \\
\hline \hline
$A(B^+ \to K^+K^0\ok)$ & 0 & $-\frac{1}{\sqrt{5}}$ & 0 & $-\frac{1}{\sqrt{5}}$ & $-\frac{1}{\sqrt{15}}$ & $-\frac{2}{3\st}$ & $\frac{3}{5}$ & 0 & $-\frac{12\s}{5\sqrt{7}}$ \\
$\frac{1}{\s} A(B^+ \to K^+K^+K^-)$ & 0 & $\frac{1}{\sqrt{5}}$ & 0 & $\frac{1}{\sqrt{5}}$ & $\frac{1}{\sqrt{15}}$ & $-\frac{1}{3\st}$ & $-\frac{3}{5}$ & 0 & $-\frac{3\s}{5\sqrt{7}}$ \\
$A(\bd \to K^0K^+K^-)$ & 0 & $\frac{1}{\sqrt{5}}$ & 0 & $\frac{1}{\sqrt{5}}$ & $-\frac{1}{\sqrt{15}}$ & $-\frac{2}{3\st}$ & $\frac{1}{5}$ & $-\frac{\s}{3}$ & $-\frac{9\s}{5\sqrt{7}}$ \\
$\frac{1}{\s} A(\bd \to K^0K^0\ok)$ & 0 & $-\frac{1}{\sqrt{5}}$ & 0 & $-\frac{1}{\sqrt{5}}$ & $\frac{1}{\sqrt{15}}$ & $-\frac{1}{3\st}$ & $-\frac{1}{5}$ & $\frac{\s}{3}$ & $-\frac{6\s}{5\sqrt{7}}$ \\
\hline \hline
$\s A(\bs \to \pi^0 K^+K^-)$ & $\frac{\st}{2\sqrt{10}}$ & 0 & $\frac{\st}{2\sqrt{10}}$ & 0 & $\frac{2}{\sqrt{15}}$ & $-\frac{1}{6\st}$ & $\frac{4}{5}$ & $\frac{1}{3\s}$ & $\frac{51}{10\sqrt{14}}$ \\
$\s A(\bs \to \pi^0 K^0\ok)$ & $\frac{\st}{2\sqrt{10}}$ & 0 & $\frac{\st}{2\sqrt{10}}$ & 0 & $-\frac{2}{\sqrt{15}}$ & $\frac{1}{6\st}$ & $-\frac{4}{5}$ & $-\frac{1}{3\s}$ & $-\frac{3\sqrt{7}}{10\s}$ \\
$A(\bs \to \pi^- K^+\ok)$ & $-\frac{\st}{2\sqrt{10}}$ & 0 & $-\frac{\st}{2\sqrt{10}}$ & 0 & 0 & $-\frac{1}{2\st}$ & 0 & $-\frac{1}{\s}$ & $-\frac{3}{2\sqrt{14}}$ \\
$A(\bs \to \pi^+ K^-K^0)$ & $-\frac{\st}{2\sqrt{10}}$ & 0 & $-\frac{\st}{2\sqrt{10}}$ & 0 & 0 & $\frac{1}{2\st}$ & 0 & $\frac{1}{\s}$ & $-\frac{3}{2\sqrt{14}}$ \\
\hline\hline
$\frac{2}{\st} A(\bs \to \pi^0\pi^0\pi^0)$ & 0 & 0 & 0 & 0 & $-\frac{2}{\sqrt{15}}$ & $-\frac{1}{3\st}$ & $-\frac{4}{5}$ & $\frac{\s}{3}$ & $\frac{6\s}{5\sqrt{7}}$ \\
$\s A(\bs \to \pi^0\pi^+\pi^-)$ & 0 & 0 & 0 & 0 & $\frac{2}{\sqrt{15}}$ & $\frac{1}{3\st}$ & $\frac{4}{5}$ & $-\frac{\s}{3}$ & $-\frac{6\s}{5\sqrt{7}}$ \\
\hline\hline
\end{tabular}
\end{center}
\end{table}

\begin{table}
\renewcommand*{\arraystretch}{2}
\caption{Amplitudes for $\Delta S = 0$ $B$-meson decays to
  fully-symmetric $PPP$ states as a function of the SU(3) matrix
  elements.
\label{tab:2}}
\begin{center}
\begin{tabular}{|c|c c|c c c c c c c|} \hline \hline
Decay & \multicolumn{2}{c|}{$V^*_{cb}V^{}_{cd}$} & \multicolumn{7}{c|}{$V^*_{ub}V^{}_{ud}$} \\ \cline{2-10}
Amplitude & $B^{\FS}_1$ & $B^{\FS}$ & $A^{\FS}_1$ & $A^{\FS}$ & $R^{\FS}_8$ & $R^{\FS}_{10}$ & $P^{\FS}_8$ & $P^{\FS}_{10^*}$ & $P^{\FS}_{27}$ \\ \hline \hline
$A(B^+ \to \pi^+K^0\ok)$ & 0 & $-\frac{1}{\sqrt{5}}$ & 0 & $-\frac{1}{\sqrt{5}}$ & $-\frac{1}{\sqrt{15}}$ & $-\frac{2}{3\st}$ & $\frac{3}{5}$ & 0 & $-\frac{12\s}{5\sqrt{7}}$ \\
$A(B^+ \to \pi^+K^+K^-)$ & 0 & $\frac{1}{\sqrt{5}}$ & 0 & $\frac{1}{\sqrt{5}}$ & $\frac{1}{\sqrt{15}}$ & $-\frac{1}{3\st}$ & $-\frac{3}{5}$ & 0 & $-\frac{3\s}{5\sqrt{7}}$ \\
$\s A(B^+ \to \pi^0K^+\ok)$ & 0 & 0 & 0 & 0 & 0 & $\frac{1}{\st}$ & 0 & 0 & $-\frac{3\s}{\sqrt{7}}$ \\
$\s A(\bd \to \pi^0K^0\ok)$ & $\frac{\st}{2\sqrt{10}}$ & $-\frac{1}{\sqrt{5}}$ & $\frac{\st}{2\sqrt{10}}$ & $-\frac{1}{\sqrt{5}}$ & $-\frac{1}{\sqrt{15}}$ & $-\frac{1}{6\st}$ & $-1$ & $\frac{1}{3\s}$ & $-\frac{9}{2\sqrt{14}}$ \\
$\s A(\bd \to \pi^0K^+K^-)$ & $\frac{\st}{2\sqrt{10}}$ & $\frac{1}{\sqrt{5}}$ & $\frac{\st}{2\sqrt{10}}$ & $\frac{1}{\sqrt{5}}$ & $\frac{1}{\sqrt{15}}$ & $\frac{1}{6\st}$ & $1$ & $-\frac{1}{3\s}$ & $-\frac{9}{2\sqrt{14}}$ \\
$A(\bd \to \pi^+K^0K^-)$ & $-\frac{\st}{2\sqrt{10}}$ & 0 & $-\frac{\st}{2\sqrt{10}}$ & 0 & 0 & $-\frac{1}{2\st}$ & 0 & $-\frac{1}{\s}$ & $-\frac{3}{2\sqrt{14}}$ \\
$A(\bd \to \pi^-K^+\ok)$ & $-\frac{\st}{2\sqrt{10}}$ & 0 & $-\frac{\st}{2\sqrt{10}}$ & 0 & 0 & $\frac{1}{2\st}$ & 0 & $\frac{1}{\s}$ & $-\frac{3}{2\sqrt{14}}$ \\
\hline \hline
$\s A(B^+ \to \pi^+\pi^0\pi^0)$ & 0 & $-\frac{1}{\sqrt{5}}$ & 0 & $-\frac{1}{\sqrt{5}}$ & $-\frac{1}{\sqrt{15}}$ & $\frac{1}{3\st}$ & $\frac{3}{5}$ & 0 & $\frac{3\s}{5\sqrt{7}}$ \\
$\frac{1}{\s} A(B^+ \to \pi^+\pi^+\pi^-)$ & 0 & $\frac{1}{\sqrt{5}}$ & 0 & $\frac{1}{\sqrt{5}}$ & $\frac{1}{\sqrt{15}}$ & $-\frac{1}{3\st}$ & $-\frac{3}{5}$ & 0 & $-\frac{3\s}{5\sqrt{7}}$ \\
$\frac{2}{\st} A(\bd \to \pi^0\pi^0\pi^0)$ & 0 & $-\frac{1}{\sqrt{5}}$ & 0 & $-\frac{1}{\sqrt{5}}$ & $-\frac{1}{\sqrt{15}}$ & $\frac{1}{3\st}$ & $-1$ & $-\frac{\s}{3}$ & 0 \\
$\s A(\bd \to \pi^0\pi^+\pi^-)$ & 0 & $\frac{1}{\sqrt{5}}$ & 0 & $\frac{1}{\sqrt{5}}$ & $\frac{1}{\sqrt{15}}$ & $-\frac{1}{3\st}$ & $1$ & $\frac{\s}{3}$ & 0 \\
\hline\hline
$\s A(\bs \to \ok\pi^0\pi^0)$ & 0 & $-\frac{1}{\sqrt{5}}$ & 0 & $-\frac{1}{\sqrt{5}}$ & $\frac{1}{\sqrt{15}}$ & $\frac{2}{3\st}$ & $-\frac{1}{5}$ & $-\frac{2\s}{3}$ & $-\frac{6\s}{5\sqrt{7}}$ \\
$A(\bs \to \ok\pi^+\pi^-)$ & 0 & $\frac{1}{\sqrt{5}}$ & 0 & $\frac{1}{\sqrt{5}}$ & $-\frac{1}{\sqrt{15}}$ & $-\frac{2}{3\st}$ & $\frac{1}{5}$ & $-\frac{\s}{3}$ & $-\frac{9\s}{5\sqrt{7}}$ \\
$\s A(\bs \to K^-\pi^+\pi^0)$ & 0 & 0 & 0 & 0 & 0 & 0 & 0 & $\s$ & $\frac{3\s}{\sqrt{7}}$ \\
\hline \hline
$\frac{1}{\s} A(\bs \to \ok K^0\ok)$ & 0 & $-\frac{1}{\sqrt{5}}$ & 0 & $-\frac{1}{\sqrt{5}}$ & $\frac{1}{\sqrt{15}}$ & $-\frac{1}{3\st}$ & $-\frac{1}{5}$ & $\frac{\s}{3}$ & $-\frac{6\s}{5\sqrt{7}}$ \\
$A(\bs \to \ok K^+K^-)$ & 0 & $\frac{1}{\sqrt{5}}$ & 0 & $\frac{1}{\sqrt{5}}$ & $-\frac{1}{\sqrt{15}}$ & $\frac{1}{3\st}$ & $\frac{1}{5}$ & $\frac{2\s}{3}$ & $-\frac{9\s}{5\sqrt{7}}$ \\
\hline\hline
\end{tabular}
\end{center}
\end{table}

Note that there are only seven combinations of matrix elements in the
amplitudes since $B^{\FS}$ and $A^{\FS}$, as well as $B^{\FS}_1$ and
$A^{\FS}_1$, always appear together:
\beq
\lambda_c^{(q)} B^{\FS} + \lambda_u^{(q)} A^{\FS} ~~,~~~~
\lambda_c^{(q)} B^{\FS}_1 + \lambda_u^{(q)} A^{\FS}_1 ~,
\label{ABcombs}
\eeq
with $\lambda_p^{(q)}=V^*_{pb} V^{}_{pq}$; $q=d,s$; $p=u,c$.

Given that there are 32 decay amplitudes, all expressed in terms of
seven combinations of matrix elements, there must be 25 independent
relations among the amplitudes. What are they?

This question is addressed as follows. The 16 $\btos$ decay amplitudes
are expressed in terms of the seven combinations of matrix elements,
so there must be nine relations among the amplitudes. Some of the
amplitude relations are determined by considering isospin alone. The
remaining relations can be found by expanding the symmetry to full
SU(3). This procedure is then applied to the 16 $\btod$ decays, giving
an additional nine amplitude relations. Finally, one can relate
certain $\btos$ and $\btod$ decay amplitudes using U spin.  There are
eight such relations, of which seven are independent of the above 18
relations. This makes a total of 25 independent relations.

\subsection{\boldmath $\btos$ Decays}

The 16 $\btos$ decays (see Table \ref{tab:1}) include $B\to K\pi\pi$
(6 decays), $B\to KK{\bar K}$ (4 decays), $\bs\to \pi K{\bar K}$ (4
decays), and $\bs\to \pi\pi\pi$ (2 decays). We begin by applying
isospin alone. All initial- and final-state particles in the decays
are eigenstates of isospin, and the weak Hamiltonian has $\Delta I =
0$ or 1. One can relate the individual amplitudes using the
Wigner-Eckart theorem.

The isospin relations among the $\btos$ $B\to PPP$ amplitudes are
(some of these are given in Ref.~\cite{3body1}):
\begin{enumerate}

\item $B\to K\pi\pi$:
\bea
{\cal A}(B^+\to K^0\pi^+\pi^0)_{\rm FS} &=& - {\cal A}(B^0_d\to K^+\pi^0\pi^-)_{\rm FS} ~, \label{iso1} \\
\s{\cal A}(B^+\to K^0\pi^+\pi^0)_{\rm FS} &=& {\cal A}(B^0_d\to K^0\pi^+\pi^-)_{\rm FS} + \s {\cal A}(B^0_d\to K^0\pi^0\pi^0)_{\rm FS} ~, \nl
\s{\cal A}(B^0_d\to K^+\pi^0\pi^-)_{\rm FS} &=& {\cal A}(B^+\to K^+\pi^+\pi^-)_{\rm FS} + \s {\cal A}(B^+\to K^+\pi^0\pi^0)_{\rm FS} ~, \nn
\eea

\item $B\to KK{\bar K}$:
\bea
{\cal A}(B^+\to K^+K^+K^-)_{\rm FS} + \s{\cal A}(B^+\to K^+K^0\ok)_{\rm FS} &=& \hskip2cm \nl
&&\hskip-7cm \s{\cal A}(B^0_d\to K^0K^+K^-)_{\rm FS} + {\cal A}(B^0_d\to K^0K^0\ok)_{\rm FS} ~,
\label{isoKKKreln}
\eea

\item $\bs\to \pi K{\bar K}$:
\bea
\s{\cal A}(B^0_s\to \pi^0 K^+K^-)_{\rm FS} + \s{\cal A}(B^0_s\to \pi^0 K^0\ok)_{\rm FS} && \nl
&& \hskip-8truecm +~ {\cal A}(B^0_s\to \pi^- K^+\ok)_{\rm FS} + {\cal A}(B^0_s\to \pi^+ K^-K^0)_{\rm FS} ~=~ 0 ~.
\eea

\item $\bs\to \pi\pi\pi$:
\bea
\frac{2}{\sqrt{3}} {\cal A}(\bs \to \piz\piz\piz)_{\rm FS} = - \sqrt{2} {\cal A}(\bs \to \piz\pip\pim)_{\rm FS} ~.
\label{iso2}
\eea

\end{enumerate}
This makes a total of six isospin relations.

Now, under isospin, the $K\pi\pi$ and $KK{\bar K}$ matrix elements
(for example) are unrelated. However, they are equal under full SU(3).
Indeed, under SU(3) all $\btos$ matrix elements are equal to their
corresponding $\btod$ matrix elements.  When one applies full SU(3),
one additional relation is
\beq
\sqrt{2} {\cal A}(B^+\to K^+\pi^+\pi^-)_{\rm FS} = {\cal A}(B^+\to K^+K^+K^-)_{\rm FS} ~.
\label{KpipiKKKrel}
\eeq
In fact, in Ref.~\cite{GRBKKK}, Gronau and Rosner showed that U spin
implies that
\beq
A(K^+\pi^+\pi^-)_{p_1 p_2 p_3} + A(K^+\pi^+\pi^-)_{p_2 p_1 p_3} = A(K^+K^+K^-)_{p_1 p_2 p_3} ~.
\label{GRrel}
\eeq
If one applies this to the fully-symmetric final state, the two terms
on the left-hand side are equal. If one also takes into account the
different normalizations for $(K^+\pi^+\pi^-)_{\rm FS}$ and
$(K^+K^+K^-)_{\rm FS}$, one gets
\bea
\frac{2}{\sqrt{6}} {\cal A}(B^+\to K^+\pi^+\pi^-)_{\rm FS} &=& \frac{1}{\sqrt{3}} {\cal A}(B^+\to K^+K^+K^-)_{\rm FS} \nn\\
\Longrightarrow \sqrt{2} {\cal A}(B^+\to K^+\pi^+\pi^-)_{\rm FS} &=& {\cal A}(B^+\to K^+K^+K^-)_{\rm FS} ~,
\eea
as above. Thus, one does not really need full SU(3) to obtain
Eq.~(\ref{KpipiKKKrel}) -- only the U-spin SU(2) subgroup is needed.

The remaining two relations are not associated with any symmetry. They
involve a number of amplitudes associated with different decays.  As
they are not particularly interesting, we do not present them here.

\subsection{\boldmath $\btod$ Decays}

The 16 $\btod$ decays (see Table \ref{tab:2}) include $B\to \pi K{\bar
  K}$ (7 decays), $B\to \pi\pi\pi$ (4 decays), $\bs\to K\pi\pi$ (3
decays), and $\bs\to KK{\bar K}$ (2 decays). Again, all initial- and
final-state particles in the decays are eigenstates of isospin, and
the weak Hamiltonian has $\Delta I = \frac12$ or $\frac32$. Using the
Wigner-Eckart theorem, the isospin relations among the $\btod$ $B\to
PPP$ amplitudes are (some of these are given in Ref.~\cite{3body1}):
\begin{enumerate}

\item $B\to \pi K{\bar K}$:
\bea
&& \sqrt{2} A(\bd \to \pi^0 K^+K^-)_{\rm FS}
+ A(\bd\to \pi^+ K^0K^-)_{\rm FS}
- A(B^+ \to \pi^+ K^+K^-)_{\rm FS} \nn\\
&& \hskip 2truecm +~\sqrt{2} A(\bd \to \pi^0 K^0\kbar)_{\rm FS}
+ A(\bd \to \pi^- K^+\kbar)_{\rm FS} \nn\\
&& \hskip 2truecm -~A(B^+ \to \pi^+ K^0\kbar)_{\rm FS}
- \sqrt{2} A(B^+ \to \pi^0 K^+\kbar)_{\rm FS} = 0 ~.
\eea

\item $B\to \pi\pi\pi$:
\bea
\sqrt{2} A(\bd\to \pi^0\pi^0\pi^0)_{\rm FS} &=& - \sqrt{3} A(\bd\to \pi^+\pi^0\pi^-)_{\rm FS} ~, \nn\\
2 A(B^+\to \pi^+\pi^0\pi^0)_{\rm FS} &=& -A(B^+\to \pi^-\pi^+\pi^+)_{\rm FS} ~.
\label{iso3}
\eea

\item $\bs\to K\pi\pi$:
\bea
\s{\cal A}(B^0_s\to \ok\pi^0\pi^0)_{\rm FS} + {\cal A}(B^0_s\to \ok\pi^+\pi^-)_{\rm FS} + \s{\cal A}(B^0_s\to K^-\pi^+\pi^0)_{\rm FS} = 0 ~.
\eea

\end{enumerate}
There are no relations involving $\bs\to KK{\bar K}$ amplitudes.  This
makes a total of four isospin relations.

As with $\btos$ decays, one can find the additional five relations by
applying full SU(3), in which case the $\btod$ matrix elements are the
same for all decays.  One of these relations,
\bea
\label{pipipi}
{\cal A}(B^+\to \pi^+\pi^+\pi^-)_{\rm FS} + \s{\cal A}(B^+\to \pi^+K^0\ok)_{\rm FS} &=& \hskip2cm \nl
&&\hskip-7cm \s{\cal A}(B^0_s\to \ok\pi^+\pi^-)_{\rm FS} + {\cal A}(B^0_s\to K^0\ok\ok)_{\rm FS} ~,
\eea
follows by applying U-spin reflection ($d \leftrightarrow s$) to the
$\btos$ relation of Eq.~(\ref{isoKKKreln}).  Another relation,
\bea
\sqrt{2} {\cal A}(B^+\to \pi^+K^+K^-)_{\rm FS} &=& {\cal A}(B^+\to \pi^+\pi^+\pi^-)_{\rm FS} ~,
\label{KKpipipipirel}
\eea
is due to full U spin~\cite{GRBKKK}, as in Eq.~(\ref{KpipiKKKrel}).

As was the case for $\btos$ decays, the remaining three relations are
not associated with any symmetry. They involve numerous amplitudes
associated with different decays.  As they are not particularly
interesting, we do not present them here.

\subsection{U Spin}
\label{Sec:Uspin}

For each $\btos$ decay in Table \ref{tab:1} whose final state does not
involve any $\pi^0$'s, there is a corresponding $\btod$ decay in Table
\ref{tab:2} related by U-spin reflection ($d \leftrightarrow s$).  The
eight pairs are
\begin{enumerate}

\item $A(\bd \to K^0 K^0 \kbar)$ and $A(\bs \to \kbar \kbar K^0)$,

\item $A(\bd \to K^0 \pi^+ \pi^-)$ and $A(\bs \to K^+ K^- \kbar)$,

\item $A(\bd \to K^+ K^0 K^-)$ and $A(\bs \to \kbar \pi^+\pi^-)$,
	
\item $A(\bs \to \pi^- K^+\kbar)$ and $A(\bd \to \pi^+ K^0 K^-)$,
	
\item $A(\bs\to \pi^+ K^0K^-)$ and $A(\bd \to \pi^- K^+ \kbar)$,
	
\item $A(B^+ \to K^+ K^+ K^-)$ and $A(B^+ \to \pi^+ \pi^+ \pi^-)$,
	
\item $A(B^+ \to K^+ \pi^+ \pi^-)$ and $A(B^+ \to \pi^+ K^+ K^-)$,

\item $A(B^+ \to K^+ K^0 \kbar)$ and $A(B^+ \to \pi^+ K^0 \kbar)$.
	
\end{enumerate}
The first (second) decay is $\btos$ ($\btod$). This shows that there
is a U-spin dependence between Eqs.~(\ref{isoKKKreln}) and
(\ref{pipipi}).

In each pair of amplitudes, terms multiplying $V^*_{cb}V^{}_{cs}$ and
$V^*_{ub}V^{}_{us}$ in one process equal terms multiplying
$V^*_{cb}V^{}_{cd}$ and $V^*_{ub}V^{}_{ud}$ in the other
\cite{Gronau:2000zy}.  (See Tables \ref{tab:1} and \ref{tab:2}.) Thus,
one can write relations among the $\btos$ and $\btod$ decay amplitudes
involving CKM matrix elements. However, it was shown in
Refs.~\cite{GRBKKK,Gronau:2000zy,Uspinbreak} that a more
experimentally-useful relation between U-spin pairs is the following:
\beq
\frac{A_s}{A_d} \, \frac{B_s}{B_d} = -1 ~,
\label{Uspinobs}
\eeq
where
\bea
B_d &=& |A(\bar b \to \bar d)|^2+|A(b \to d)|^2 ~,\nn\\
B_s &=& |A(\bar b \to \bar s)|^2+|A(b \to s)|^2 ~,\nn\\
A_d &=& \frac{|A(\bar b \to \bar d)|^2-|A(b \to d)|^2}{|A(\bar b \to \bar d)|^2+|A(b \to d)|^2}~,\nn\\
A_s &=& \frac{|A(\bar b \to \bar s)|^2-|A(b \to s)|^2}{|A(\bar b \to \bar s)|^2+|A(b \to s)|^2}~.
\eea
$B_d$ and $B_s$ are related to the CP-averaged $\btod$ and $\btos$
decay rates, while $A_d$ and $A_s$ are direct CP asymmetries. The
CP-conjugate amplitude $A(\bar b \to \bar q)$ is obtained from $A(b
\to q)$ by changing the signs of the weak phases.  (Since we have
U-spin reflections, these relations hold for all final symmetry
states.)

There are eight U-spin relations of this kind. Along with the nine
$\btos$ and nine $\btod$ decay amplitude relations, of which one pair
[Eqs.~(\ref{isoKKKreln}) and (\ref{pipipi})] is related by U-spin
reflection, this makes a total of 25 independent relations.  This is
consistent with 32 decay amplitudes all expressed as a function of
seven combinations of SU(3) matrix elements.

\section{AMPLITUDES AND DIAGRAMS}

\subsection{\boldmath $B\to PP$ decays}

In Ref.~\cite{GHLR1}, it was argued that two-body $B \to PP$ decays
can be described by six diagrams: the color-favored and
color-suppressed tree amplitudes $T$ and $C$, the gluonic-penguin
amplitude $P$, the exchange amplitude $E$, the annihilation amplitude
$A$, and the penguin-annihilation amplitude $PA$. Now, $P$ receives
contributions from the internal quarks $t$, $c$ and $u$:
\bea
P & = & V^*_{tb}V^{}_{td} P_t + V^*_{cb}V^{}_{cd} P_c + V^*_{ub}V^{}_{ud} P_u \nn\\
  & = & V^*_{cb}V^{}_{cd} P_{ct} + V^*_{ub}V^{}_{ud} P_{ut} ~,
\label{Pdef}
\eea
where $P_{ct} \equiv P_c - P_t$, $P_{ut} \equiv P_u - P_t$, and the
unitarity of the CKM matrix [Eq.~(\ref{unitarity})] has been used in
the second line.  Note that two-body diagrams are generally
(re)defined to absorb the CKM matrix elements that multiply them.
That is, $T$, $P_{ut}$ , etc.\ include $V^*_{ub}V^{}_{ud}$, and $P_{ct}$
and $PA_{ct}$ include $V^*_{cb}V^{}_{cd}$.

In $B\to PP$ decays there are seven matrix elements. $\{1\}_u$,
$\{8_1\}_u$, $\{8_2\}$, $\{8_3\}$, and $\{27\}$ are all multiplied by
$V^*_{ub}V^{}_{ud}$, while $\{1\}_c$ and $\{8_1\}_c$ are multiplied by
$V^*_{cb}V^{}_{cd}$. For $\Delta C = 0,~\Delta S = 0$ decays, the
following relations between matrix elements and diagrams were
established \cite{GHLR1}:
\bea
 V^*_{ub}V^{}_{ud} \, \{1\}_u &=& 2\st \left[ PA_{ut} + \frac{2}{3}P_{ut} + \frac{2}{3}E - \frac{1}{12}C
+ \frac{1}{4}T \right] ~,  \nn\\
 V^*_{ub}V^{}_{ud} \, \{8_1\}_u &=& - \sqrt{\frac{5}{3}} \left[ P_{ut} + \frac{3}{8}(T+A) - \frac{1}{8}
(C + E) \right] ~,  \nn\\
 V^*_{ub}V^{}_{ud} \, \{8_2\} &=& \frac{\sqrt{5}}{4}(C + A - T - E ) ~,  \nn\\
 V^*_{ub}V^{}_{ud} \, \{8_3\} &=& - \frac{1}{8\sqrt{3}}(T + C ) - \frac{5}{8\sqrt{3}}(A + E) ~,  \nn\\
 V^*_{ub}V^{}_{ud} \, \{27\} &=& -\frac{1}{2\st} \left[T + C \right] ~,  \nn\\
 V^*_{cb}V^{}_{cd} \, \{1\}_c &=& 2\st \left[ PA_{ct} + \frac{2}{3}P_{ct} \right] ~,  \nn\\
 V^*_{cb}V^{}_{cd} \, \{8_1\}_c &=& - \sqrt{\frac{5}{3}} P_{ct} ~.
\label{PPrels}
\eea
A corresponding set of relations exists for the diagrams corresponding
to strangeness-changing $\Delta C = 0$ decays. This shows that the
description of decay amplitudes in terms of diagrams is equivalent to
an SU(3) description.

Above, the equivalence of diagrams and SU(3) is shown for $B \to PP$
decays. However, there are two diagrams that have not been included:
the color-favored and color-suppressed electroweak-penguin (EWP)
amplitudes $\pew$ and $\pewc$ \cite{GHLR2}. If they are added, the
relations between matrix elements and diagrams are modified, but this
does not change the fact that diagrams and SU(3) are equivalent.
Indeed, the addition of EWP's only has the effect of redefining the
diagrams:
\bea
T & \to & T + \pewc ~, \nn\\
C & \to & C + \pew ~, \nn\\
(P_{ut} + P_{ct}) & \to & (P_{ut} + P_{ct}) - \frac13 \pewc ~.
\eea
Although this shows that, in fact, EWP's are not independent, it does
not indicate how EWP's are related to the other diagrams. The actual
$B \to PP$ EWP-tree relations were found later, in
Refs.~\cite{NR,GPY}.

\subsection{\boldmath $B\to PPP$ decays}

In Ref.~\cite{3body1}, it was argued that $B \to PPP$ decays can also
be described by diagrams, similar to those of $B \to PP$. For the
three-body analogues of $T$, $C$ and $P$, one has to ``pop'' a quark
pair from the vacuum.  The subscript ``1'' (``2'') is added if the
popped quark pair is between two non-spectator final-state quarks (two
final-state quarks including the spectator). For the $P$-type
diagrams, it turns out that only the combination ${\tilde P} \equiv
P_1 + P_2$ appears in amplitudes. For the three-body analogues of $E$,
$A$ and $PA$, the spectator quark interacts with the ${\bar b}$, and
one has two popped quark pairs. Here there is only one of each type of
diagram, so there are a total of eight diagrams: $T_{1,2}$, $C_{1,2}$,
${\tilde P}$, $E$, $A$, $PA$. Furthermore, for each of ${\tilde P}$
and $PA$, we allow for two contributions, giving ${\tilde P}_{ut}$,
${\tilde P}_{ct}$, $PA_{ut}$ and $PA_{ct}$, where ${\tilde P}_{ut}
\equiv {\tilde P}_u - {\tilde P}_t$, and similarly for the other
diagrams. For certain decays, the diagrams may have a popped $s{\bar
  s}$ quark pair. Under SU(3), these are equal to the same diagrams
with a popped $d{\bar d}$ or $u{\bar u}$. As in $B\to PP$ decays, we
define the diagrams to absorb the CKM matrix elements that multiply
them.  The decomposition of all 32 amplitudes in terms of diagrams is
given in Tables \ref{tab:3} and \ref{tab:4}.

\begin{table}
\renewcommand*{\arraystretch}{2}
\caption{Amplitudes for $\Delta S = 1$ $B$-meson decays to
  fully-symmetric $PPP$ states as a function of the three-body
  diagrams ($\btos$ diagrams are written with primes).
\label{tab:3}}
\begin{center}
\begin{tabular}{|c|c c|c c c c c c c c|} \hline \hline
Decay & \multicolumn{2}{c|}{$V^*_{cb}V^{}_{cs}$} & \multicolumn{8}{c|}{$V^*_{ub}V^{}_{us}$} \\ \cline{2-11}
Amplitude & ${\tilde P}'_{ct}$ & $PA'_{ct}$ & ${\tilde P}'_{ut}$ & $C'_1$
& $C'_2$ & $T'_1$ & $T'_2$ & $E'$ & $A'$ & $PA'_{ut}$ \\ \hline \hline
$A(B^+ \to K^+\pi^+\pi^-)$ & $-1$ & 0 & $-1$ & $-1$ & 0 & 0 & $-1$ & 0 & $-1$ & 0 \\
$\s A(B^+ \to K^+\pi^0\pi^0)$ & 1 & 0 & 1 & 1 & 1 & 1 & 1 & 0 & 1 & 0 \\
$\s A(B^+ \to K^0\pi^+\pi^0)$ & 0 & 0 & 0 & 0 & $-1$ & $-1$ & 0 & 0 & 0 & 0 \\
$A(\bd \to K^0\pi^+\pi^-)$ & $-1$ & 0 & $-1$ & $-1$ & 0 & $-1$ & 0 & 0 & 0 & 0 \\
$\s A(\bd \to K^0\pi^0\pi^0)$ & 1 & 0 & 1 & 1 & $-1$ & 0 & 0 & 0 & 0 & 0 \\
$\s A(\bd \to K^+\pi^0\pi^-)$ & 0 & 0 & 0 & 0 & 1 & 1 & 0 & 0 & 0 & 0 \\
\hline \hline
$A(B^+ \to K^+K^0\ok)$ & 1 & 0 & 1 & 0 & 0 & 0 & 0 & 0 & 1 & 0 \\
$\frac{1}{\s} A(B^+ \to K^+K^+K^-)$ & $-1$ & 0 & $-1$ & $-1$ & 0 & 0 & $-1$ & 0 & $-1$ & 0 \\
$A(\bd \to K^0K^+K^-)$ & $-1$ & 0 & $-1$ & $-1$ & 0 & 0 & $-1$ & 0 & 0 & 0 \\
$\frac{1}{\s} A(\bd \to K^0K^0\ok)$ & 1 & 0 & 1 & 0 & 0 & 0 & 0 & 0 & 0 & 0 \\
\hline \hline
$\s A(\bs \to \pi^0 K^+K^-)$ & 1 & 1 & 1 & 0 & 1 & 1 & 1 & 2 & 0 & 1 \\
$\s A(\bs \to \pi^0 K^0\ok)$ & 1 & 1 & 1 & 0 & $-1$ & 0 & 0 & 0 & 0 & 1 \\
$A(\bs \to \pi^- K^+\ok)$ & $-1$ & $-1$ & $-1$ & 0 & 0 & 0 & $-1$ & $-1$ & 0 & $-1$ \\
$A(\bs \to \pi^+ K^-K^0)$ & $-1$ & $-1$ & $-1$ & 0 & 0 & $-1$ & 0 & $-1$ & 0 & $-1$ \\
\hline\hline
$\frac{2}{\st} A(\bs \to \pi^0\pi^0\pi^0)$ & 0 & 0 & 0 & 0 & 0 & 0 & 0 & $-1$ & 0 & 0 \\
$\s A(\bs \to \pi^0\pi^+\pi^-)$ & 0 & 0 & 0 & 0 & 0 & 0 & 0 & 1 & 0 & 0 \\
\hline\hline
\end{tabular}
\end{center}
\end{table}

\begin{table}
\renewcommand*{\arraystretch}{2}
\caption{Amplitudes for $\Delta S = 0$ $B$-meson decays to
  fully-symmetric $PPP$ states as a function of the three-body
  diagrams.
\label{tab:4}}
\begin{center}
\begin{tabular}{|c|c c|c c c c c c c c|} \hline \hline
Decay & \multicolumn{2}{c|}{$V^*_{cb}V^{}_{cd}$} & \multicolumn{8}{c|}{$V^*_{ub}V^{}_{ud}$} \\ \cline{2-11}
Amplitude & ${\tilde P}_{ct}$ & $PA_{ct}$ & ${\tilde P}_{ut}$ & $C_1$
& $C_2$ & $T_1$ & $T_2$ & $E$ & $A$ & $PA_{ut}$ \\ \hline \hline
$A(B^+ \to \pi^+K^0\ok)$ & 1 & 0 & 1 & 0 & 0 & 0 & 0 & 0 & 1 & 0 \\
$A(B^+ \to \pi^+K^+K^-)$ & $-1$ & 0 & $-1$ & $-1$ & 0 & 0 & $-1$ & 0 & $-1$ & 0 \\
$\sqrt{2} A(B^+ \to \pi^0K^+\ok)$ & 0 & 0 & 0 & 0 & $-1$ & $-1$ & 0 & 0 & 0 & 0 \\
$\s A(\bd \to \pi^0K^0\ok)$ & 2 & 1 & 2 & 0 & $-1$ & 0 & 0 & 0 & 0 & 1 \\
$\s A(\bd \to \pi^0K^+K^-)$ & 0 & 1 & 0 & $-1$ & 0 & 0 & 0 & 2 & 0 & 1 \\
$A(\bd \to \pi^+K^0K^-)$ & $-1$ & $-1$ & $-1$ & 0 & 0 & 0 & $-1$ & $-1$ & 0 & $-1$ \\
$A(\bd \to \pi^-K^+\ok)$ & $-1$ & $-1$ & $-1$ & 0 & 0 & $-1$ & 0 & $-1$ & 0 & $-1$ \\
\hline \hline
$\s A(B^+ \to \pi^+\pi^0\pi^0)$ & 1 & 0 & 1 & 1 & 0 & 0 & 1 & 0 & 1 & 0 \\
$\frac{1}{\s} A(B^+ \to \pi^+\pi^+\pi^-)$ & $-1$ & 0 & $-1$ & $-1$ & 0 & 0 & $-1$ & 0 & $-1$ & 0 \\
$\frac{2}{\st} A(\bd \to \pi^0\pi^0\pi^0)$ & 1 & 0 & 1 & 1 & $-1$ & 0 & 0 & $-1$ & 0 & 0 \\
$\s A(\bd \to \pi^0\pi^+\pi^-)$ & $-1$ & 0 & $-1$ & $-1$ & 1 & 0 & 0 & 1 & 0 & 0 \\
\hline\hline
$\s A(\bs \to \ok\pi^0\pi^0)$ & 1 & 0 & 1 & 1 & $-1$ & 0 & 0 & 0 & 0 & 0 \\
$A(\bs \to \ok\pi^+\pi^-)$ & $-1$ & 0 & $-1$ & $-1$ & 0 & 0 & $-1$ & 0 & 0 & 0 \\
$\s A(\bs \to K^-\pi^+\pi^0)$ & 0 & 0 & 0 & 0 & 1 & 0 & 1 & 0 & 0 & 0 \\
\hline \hline
$\frac{1}{\s} A(\bs \to \ok K^0\ok)$ & 1 & 0 & 1 & 0 & 0 & 0 & 0 & 0 & 0 & 0 \\
$A(\bs \to \ok K^+K^-)$ & $-1$ & 0 & $-1$ & $-1$ & 0 & $-1$ & 0 & 0 & 0 & 0 \\
\hline\hline
\end{tabular}
\end{center}
\end{table}

For the $\btod$ decays there are eight diagrams that include
$V^*_{ub}V^{}_{ud}$ and two that include $V^*_{cb}V^{}_{cd}$. Similarly,
there are seven matrix elements proportional to $V^*_{ub}V^{}_{ud}$ and
two proportional to $V^*_{cb}V^{}_{cd}$.  By analyzing all 16 $\btod$
decays, one finds that the relations between these are
\bea
V^*_{ub}V^{}_{ud} \, A^{\FS}_1 &=& \frac{\sqrt{5}}{2\sx} (8 {\tilde P}_{ut} + 8 PA_{ut} + 3 T_1 + 3 T_2 - C_1 - C_2 + 8 E)~, \nl
V^*_{ub}V^{}_{ud} \, A^{\FS}   &=& \frac{\sqrt{5}}{8} (-8 {\tilde P}_{ut} - T_1 - 3 T_2 - 5 C_1 + C_2 + E - 3 A) ~, \nl
V^*_{ub}V^{}_{ud} \, R_8^{\FS} &=& \frac{\sqrt{5}}{4\st} (T_1 - T_2 - C_1 + C_2 + 3 E - 3 A) ~, \nl
V^*_{ub}V^{}_{ud} \, R_{10}^{\FS} &=& \frac{\st}{2} (-T_1 + T_2 + C_1 - C_2) ~, \nl
V^*_{ub}V^{}_{ud} \, P_8^{\FS} &=& \frac{1}{24}(T_1 + 11 T_2 + C_1 + 11 C_2 + 15 E + 15 A) ~, \nl
V^*_{ub}V^{}_{ud} \, P_{10^*}^{\FS} &=& \frac{1}{2\s} (-T_1 + T_2 - C_1 + C_2) ~, \nl
V^*_{ub}V^{}_{ud} \, P_{27}^{\FS} &=& \frac{\sqrt{7}}{6\s}  (T_1 + T_2 + C_1 + C_2) ~, \nl
V^*_{cb}V^{}_{cd} \, B_1^{(fs)} &=& 2 \sqrt{\frac{10}{3}} ({\tilde P}_{ct}  +  PA_{ct}) ~, \nl
V^*_{cb}V^{}_{cd} \, B^{(fs)} &=& -\sqrt{5} {\tilde P}_{ct} ~.
\label{PPPrels}
\eea
A corresponding set of relations exists for the diagrams corresponding
to $\btos$ decays. This demonstrates the equivalence of diagrams and
SU(3) for the fully-symmetric $PPP$ state.

Now, there are also four EWP diagrams that can contribute to the
amplitudes: $P_{EW1,2}$ and $P^C_{EW1,2}$. However, as
Eq.~(\ref{PPPrels}) shows, in $B \to PPP$ the equivalence of diagrams
and SU(3) holds without EWP's (as was the case for $B \to PP$). The
EWP diagrams are therefore not independent. Their addition only has
the effect of redefining the diagrams:
\bea
T_i & \to & T_i + P^C_{EWi} ~, \nn\\
C_i & \to & C_i + P_{EWi} ~, \nn\\
({\tilde P}_{ut} + {\tilde P}_{ct}) & \to & ({\tilde P}_{ut} + {\tilde P}_{ct}) - \frac23 P_{EW1} - \frac13 (P^C_{EW1} + P^C_{EW2}) ~, \nn\\
(PA_{ut} + PA_{ct}) & \to & (PA_{ut} + PA_{ct}) + \frac23 P_{EW1} ~,
\eea
where $i=1,2$.  The above shows that EWP contributions do not
introduce new independent SU(3) amplitudes, but does not indicate how
EWP's are related to the other diagrams. The exact EWP-tree relations
for the fully-symmetric state were given in Ref.~\cite{3body2}. Taking
$c_1/c_2 = c_9/c_{10}$ (which holds to about 5\%), the simplified form
is
\begin{equation}
P_{EWi} = \kappa T_i ~~,~~~~ P^C_{EWi} = \kappa C_i ~,
\end{equation}
where
\begin{equation}
\kappa \equiv - \frac{3}{2} \frac{|\lambda_t^{(d)}|}{|\lambda_u^{(d)}|} \frac{c_9 + c_{10}}{c_1 + c_2} ~.
\end{equation}

\subsection{\boldmath Neglect of $E$/$A$/$PA$}

The utility of expressing the $B \to PPP$ amplitudes in terms of
diagrams is that it is relatively easy to ascertain which diagrams
contribute to a given decay amplitude, while it is much more difficult
to do this for the SU(3)-matrix elements. However, the great advantage
of using diagrams is that it provides dynamical input. In particular,
the diagrams $E$, $A$ and $PA$ all involve the interaction of the
spectator quark. As such, they are expected to be considerably smaller
than the $T_i$, $C_i$ and ${\tilde P}$, and can therefore be
neglected, to a first approximation.

In $B\to PPP$ decays, if $E$/$A$/$PA$ are neglected, one has
\bea
R_{10}^{\FS} &=& -\frac{6}{\sqrt{5}} R_8^{\FS}~,\nn\\
P_{27}^{\FS} &=& \frac{1}{18} \sqrt{\frac{7}{2}} (12 P_8^{\FS} - 5\sqrt{2}
 P_{10^*}^{\FS})~, \nn\\
B_1^{\FS} &=& - 2\sqrt\frac{2}{3} B^{\FS}~.
\eea
The first two relations above reduce the number of combinations of
reduced matrix elements in SU(3) from seven to five. However, because
$B_1^{\FS}$ and $B^{\FS}$ always appear with $A^{\FS}_1$ and
$A^{\FS}$, respectively [Eq.~(\ref{ABcombs})], the third relation does
not lead to a further change in the number of combinations of reduced
matrix elements.  Given that the 16 $\btos$ decay amplitudes are now
expressed in terms of five combinations of matrix elements, there must
be 11 independent relations among the amplitudes. That is, there must
be two additional relations, and similarly for the $\btod$
amplitudes. What are they?

For $\btos$, the additional relations are as follows:
\begin{enumerate}

\item The $KK{\bar K}$ relation of Eq.~(\ref{isoKKKreln}) is split
  into two relations:
\bea
A(B^+ \to K^+ K^+ K^-)_{\rm FS} &=& \sqrt{2} A(\bd \to K^+ K^0 K^-)_{\rm FS} ~, \nn\\
\sqrt{2} A(B^+ \to K^+ K^0 \kbar)_{\rm FS} &=& A(\bd \to K^0 K^0 \kbar)_{\rm FS} ~.
\eea

\item ${\cal A}(\bs \to \piz\pip\pim)_{\rm FS} = 0$, i.e., the
two decays $\bs \to \piz\pip\pim$ and $\bs\to\piz\piz\piz$ are pure $E'$.

\end{enumerate}
For $\btod$, the additional relations are as follows:
\begin{enumerate}

\item $A(\bs \to \ok\pi^+\pi^-)_{\rm FS} = A(B^+ \to \pi^+K^+K^-)_{\rm FS}$.

\item $A(B^+ \to \pi^+\pi^0\pi^0)_{\rm FS} + A(\bd \to
  \pi^0\pi^+\pi^-)_{\rm FS} = A(\bs \to K^-\pi^+\pi^0)_{\rm FS}$.

\end{enumerate}
The above relations are not due to group theory alone -- there is
dynamical input.

The remaining five relations among the $B \to PPP$ amplitudes can be
found among the eight U-spin relations described in Sec.~IIIC. In
fact, for the case where $E$/$A$/$PA$ are neglected, most of the $B
\to PPP$ amplitude relations can be cast in terms of U-spin relations.
The full list, including the true U-spin pairs, is \cite{Uspinbreak}
\begin{enumerate}
	
\item ($\bd \to K^+K^-K^0$, $B^+ \to K^+K^+K^-$, $B^+ \to K^+
  \pi^+\pi^-$) and ($B^+ \to \pi^+ K^- K^+$, $B^+ \to \pi^+ \pi^0
  \pi^0$, $B^+ \to \pi^+ \pi^+ \pi^-$, $\bs \to \kbar \pi^+\pi^-$),
	
\item ($B^+ \to K^+ K^0 \kbar$, $\bd \to K^0 K^0 \kbar$) and ($B^+ \to
  \pi^+ K^0 \kbar$, $\bs \to \kbar \kbar K^0$),
	
\item $\bd \to K^0\pi^+\pi^-$ and $\bs \to \ok K^+K^-$,

\item $\bs \to \pi^- K^+\ok$ and $\bd \to \pi^+K^0K^-$,

\item $\bs \to \pi^+ K^-K^0$ and $\bd \to \pi^-K^+\ok$,

\item ($B^+ \to K^0 \pi^+ \pi^0$, $\bd \to K^+ \pi^- \pi^0$) and $B^+
  \to K^+ \kbar \pi^0$,
	
\item $\bd \to K^0 \pi^0 \pi^0$ and ($\bd \to \pi^0 \pi^0 \pi^0$, $\bd
  \to \pi^0\pi^+\pi^-$, $\bs \to \ok\pi^0\pi^0$).

\end{enumerate}
The decays in the first (second) parentheses are $\btos$ ($\btod$)
transitions.

\section{EXPERIMENTAL TESTS}

We have expressed all decay amplitudes in terms of matrix elements,
and shown the equivalence with the diagrammatic description. However,
the real goal is to produce predictions that can be tested
experimentally.  Some of the amplitude relations we have presented
relate more than two decay processes and hence involve relative strong
phases among amplitudes. Since such strong phases are difficult to
measure, these relations are hard to verify experimentally. Thus, the
most interesting of them are equalities between two amplitudes.
Although there can be a relative strong phase between the two
amplitudes, such a phase does not affect the relationship between the
magnitudes, which are much easier to measure. In this section we focus
on such ``two-amplitude equalities'' and discuss how they can be
tested.

\subsection{\boldmath Measuring the $PPP$ fully-symmetric amplitude}

When considering any experimental tests, the first question is: how
can the fully-symmetric final state be probed? The method for doing
this was discussed in Ref.~\cite{3body1}; it proceeds as follows.

For the decay $B \to P_1 P_2 P_3$, one defines the three Mandelstam
variables $s_{ij} \equiv \left( p_i + p_j \right)^2$, where $p_i$ is
the momentum of each $P_i$. These are not independent, but obey
$s_{12} + s_{13} + s_{23} = m_B^2 + m_1^2 + m_2^2 + m_3^2$. The $B \to
P_1 P_2 P_3$ Dalitz plot is given in terms of two Mandelstam
variables, say $s_{12}$ and $s_{13}$. Now, one can obtain the decay
amplitude ${\cal M} (s_{12}, s_{13})$ describing this Dalitz plot. To
do this, one uses an isobar model. Here the amplitude is expressed as
the sum of a non-resonant and several intermediate resonant
contributions:
\beq
{\cal M} (s_{12}, s_{13}) = {\cal N}_{\rm DP}\sum\limits_j c_j e^{i \theta_j} F_j
(s_{12}, s_{13})~,
\eeq
where the index $j$ runs over all contributions. Each contribution is
expressed in terms of isobar coefficients $c_j$ (magnitude) and
$\theta_j$ (phase), and a dynamical wave function $F_j$. ${\cal
  N}_{\rm DP}$ is a normalization constant. The $F_j$ take different
forms depending on the contribution. The $c_j$ and $\theta_j$ are
extracted from a fit to the Dalitz-plot event distribution.

Now, given ${\cal M} (s_{12}, s_{13})$, one can construct the
fully-symmetric amplitude. It is given simply by
\bea \label{fulsym}
{\cal M}_{\rm FS} &=&
\frac{1}{\sqrt{6}} \left[ {\cal M}(s_{12},s_{13}) + {\cal M}(s_{13},s_{12})
+ {\cal M}(s_{12},s_{23}) \right. \nn\\
&& \hskip2truecm \left.+~{\cal M}(s_{23},s_{12}) + {\cal M}(s_{23},s_{13})
+ {\cal M}(s_{13},s_{23}) \right]~.
\eea
Thus, for any three-body decay for which a Dalitz plot has been
measured, one can extract the fully-symmetric amplitude.

\subsection{SU(3) amplitude equalities}

As noted above, two-amplitude equalities are the most interesting
since they are subject to direct experimental tests. There are four
such equalities under isospin symmetry, given in Eqs.~(\ref{iso1}),
(\ref{iso2}) and (\ref{iso3}).  These isospin relations may be useful
as a test of the SM, constraining new-physics models in which the
effective Hamiltonian involves strangeness-conserving $\Delta I=5/2$
operators or strangeness-changing $\Delta I =2$ operators.

If one restricts attention to final states involving only charged
kaons and pions, there are two equalities due to U spin; they relate
two $\btos$ decays, and two $\btod$ decays. They are given in
Eqs.~(\ref{KpipiKKKrel}) and (\ref{KKpipipipirel}), and repeated for
convenience below:
\bea
\sqrt{2} {\cal A}(B^+\to K^+\pi^+\pi^-)_{\rm FS} & = & {\cal A}(B^+\to K^+K^+K^-)_{\rm FS} ~, \nn\\
\sqrt{2} {\cal A}(B^+\to \pi^+K^+K^-)_{\rm FS} &=& {\cal A}(B^+\to \pi^+\pi^+\pi^-)_{\rm FS} ~.
\eea
Consider the first equality, the ``$K\pi\pi$-$KK{\bar K}$ relation.''
Since we are not interested in the overall phase of the amplitudes,
the relation can be written as
\beq
\frac{|{\cal A}(B^+\to K^+K^+K^-)_{\rm FS}|}{\sqrt{2} |{\cal A}(B^+\to K^+\pi^+\pi^-)_{\rm FS}|} = 1 ~,
\label{relnexpt}
\eeq
and note that it also holds for $B^-$ decays.  The key point is that
this relation holds at every point in the Dalitz plot (though we can
only use one sixth of the Dalitz plot due to the fact that the
amplitudes are fully symmetric). Thus, this ratio should be measured
for each Dalitz-plot point, and then one should average over all
points. This is extremely important, as it has the effect of reducing
the errors. (As always, correlations among the different points must
be taken into account in calculating the error.)

\begin{figure}
\begin{center}
\includegraphics[width=0.69\textwidth]{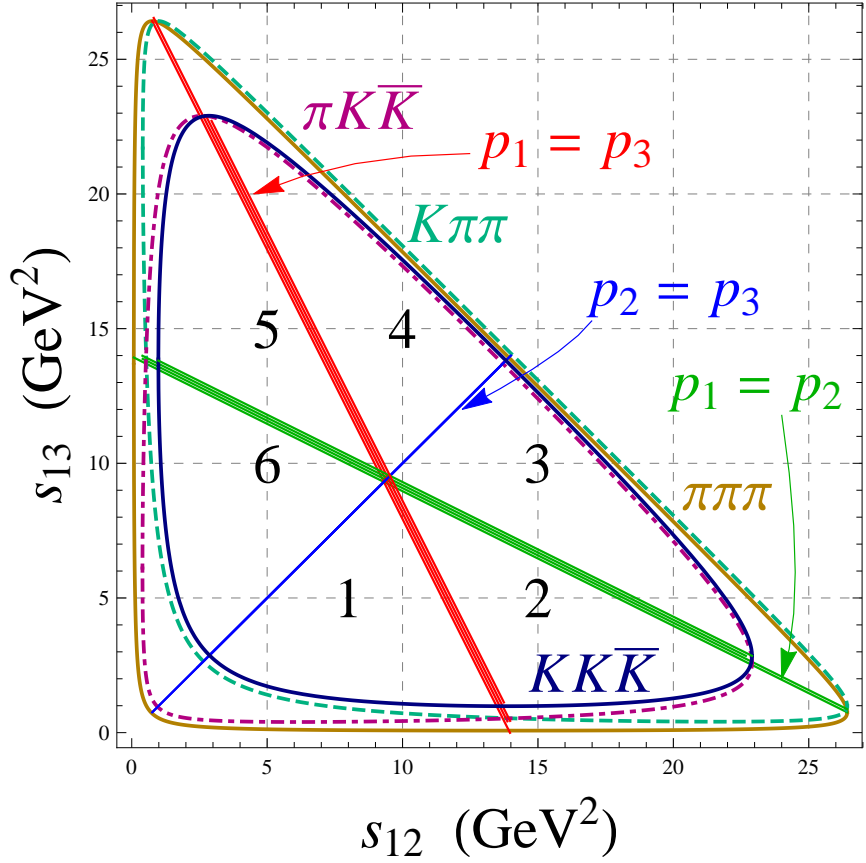}
\end{center}
\caption{Dalitz-plot regions and boundaries for $B\to PPP$ decays.
  Outer solid curve: $\pi \pi \pi$; dashed curve: $K \pi \pi$;
  dot-dashed curve: $\pi K \bar K$; inner solid curve: $K K \bar K$.
\label{fig:DP}}
\end{figure}

Of course, SU(3)-breaking effects can lead to a violation of the
$K\pi\pi$-$KK{\bar K}$ relation.  One can see such effects, for
instance, in the different boundaries of the Dalitz plots for $B \to
\pi\pi\pi,~K \pi \pi,~\pi K \bar K$, and $K K \bar K$, as shown in
Fig.\ \ref{fig:DP}.  In order to treat SU(3) breaking rigorously, one
should introduce into the $B \to PPP$ matrix elements a spurion mass
term $M_{brk} \sim (2s{\bar s} - u{\bar u} - d{\bar d})$, behaving as
an SU(3) octet. When one does this, one finds several SU(3)-breaking
terms. However, as far as the $K\pi\pi$-$KK{\bar K}$ relation is
concerned, the overall effect is to modify the two-amplitude
equality. In the presence of SU(3) breaking, we have
\beq
{\cal A}(B^+\to K^+K^+K^-)_{\rm FS} = X \sqrt{2} {\cal A}(B^+\to K^+\pi^+\pi^-)_{\rm FS} ~,
\eeq
where $X$ is the SU(3)-breaking factor. Note that $X$ does not take
the same group-theoretical form at each point in the Dalitz plot. It
is simply a complex number that affects the equality.
Eq.~(\ref{relnexpt}) is now modified:
\beq
\frac{|{\cal A}(B^+\to K^+K^+K^-)_{\rm FS}|}{\sqrt{2} |{\cal A}(B^+\to K^+\pi^+\pi^-)_{\rm FS}|} = |X| ~.
\eeq
Now, as is always the case, the size of SU(3) breaking is unknown
(though it is typically $\lsim 25\%$). It is logically possible that
$|X| - 1 > 0$ at each point in the Dalitz plot, perhaps very much
larger. However, given that $X$ depends in a complicated way on all
the SU(3)-breaking terms, it seems more likely that $|X| - 1 > 0$ at
some points, and $|X| - 1 < 0$ at others.  In this situation,
averaging over all Dalitz-plot points will {\it reduce} the effect of
SU(3) breaking.
%
Of course, there is no guarantee that this occurs for the
$K\pi\pi$-$KK{\bar K}$ relation, but whether or not it happens will be
determined experimentally.  And SU(3) breaking could be smaller simply
due to the fact that all spin-1 resonances, and the associated SU(3)
breaking, are absent from the fully-symmetric amplitudes.

Indeed, this type of effect was seen in Ref.~\cite{BIL}, where some
$\btokpipi$ and $\btokkk$ decays were analyzed. There, the
approximation was made that, to leading order, the SU(3) breaking in
all diagrams is equal, so that $A(B \to KK{\bar K}) = \alpha_{SU(3)}
A(B \to K\pi\pi)$, where $\alpha_{SU(3)}$ measures the amount of SU(3)
breaking. Averaged over the entire Dalitz plot, it was found that
$|\alpha_{SU(3)}| = 0.97 \pm 0.05$, i.e., an SU(3) breaking of $\sim
5\%$ was found. Perhaps something like this happens with the
$K\pi\pi$-$KK{\bar K}$ relation.

Now, LHCb has measured \cite{LHCbDS1}
\bea\label{asym1}
A_{CP}(B^+\to K^+\pi^+\pi^-) &=& +0.032 \pm 0.008 ({\rm stat}) \pm 0.004
({\rm syst}) \pm 0.007 (J/\psi K^+) ~, \nn\\
A_{CP}(B^+\to K^+K^+K^-) &=& -0.043 \pm 0.009 ({\rm stat}) \pm 0.003
({\rm syst}) \pm 0.007 (J/\psi K^+) ~,
\eea
and \cite{Aaij:2013bla}
\bea\label{asym2}
A_{CP}(B^+\to \pi^+K^+K^-) &=& -0.141 \pm 0.040 ({\rm stat}) \pm 0.018
({\rm syst}) \pm 0.007 (J/\psi K^+) ~, \nn\\
A_{CP}(B^+\to \pi^+\pi^+\pi^-) &=& +0.117 \pm 0.021 ({\rm stat}) \pm 0.009
({\rm syst}) \pm 0.007 (J/\psi K^+) ~,
\eea
with larger asymmetries, with the same signs as the above, observed in
localized regions of phase space. As with all measurements, one wants
to compare these results with the predictions of the SM.

Using measured values of corresponding branching ratios~\cite{PDG},
relative signs and relative magnitudes of the above asymmetries were
shown to be in modest and reasonable agreement with two U-spin
predictions~\cite{Bhattacharya:2013cvn},
\bea\label{ratio1}
\frac{A_{CP}(B^+\to \pi^+ K^+ K^-)}{A_{CP}(B^+ \to K^+ \pi^+ \pi^-)} & = &
-\frac{{\cal B}(B^+\to K^+ \pi^+\pi^-)}{{\cal B}(B^+ \to \pi^+ K^+ K^-)}~,\\
\frac{A_{CP}(B^+ \to \pi^+ \pi^+\pi^-)}{A_{CP}(B^+ \to K^+K^+K^-)} & = &
-\frac{{\cal B}(B^+ \to K^+K^+ K^-)}{{\cal B}(B^+ \to \pi^+ \pi^+ \pi^-)}~.
\label{ratio2}
\eea
On the other hand, SM predictions for the asymmetries themselves are
much less certain because direct CP asymmetries involve unknown strong
phases.

The $K\pi\pi$-$KK{\bar K}$ relation [Eq.~(\ref{relnexpt})] and a
similar $\pi K{\bar K}$-$\pi\pi\pi$ relation provide clean tests of
the SM within the flavor SU(3) approximation. They say that, for the
fully-symmetric states, the amplitudes of the two pairs of decays,
whatever values they take, are equal at all points of the Dalitz plot.
As we have argued, SU(3) breaking in these two equalities is expected
to be reduced when averaged over the entire Dalitz plot.  In order to
test these two relations, LHCb must extract the fully-symmetric
amplitudes for the four decays. As noted above, this requires
performing an isobar analysis of the Dalitz plots of $B^+\to
K^+\pi^+\pi^-$, $B^+\to K^+K^+K^-$, $B^+\to\pi^+K^+K^-$ and $B^+\to
\pi^+\pi^+\pi^-$.  However, LHCb already has these Dalitz-plot data,
so it is straightforward to test the SM using the $K\pi\pi$-$KK{\bar
  K}$ and $\pi K{\bar K}$-$\pi\pi\pi$ relations. This can be done now,
and we strongly encourage LHCb to carry out these analyses.

Finally, in Sec.~IIIC it was noted that there are eight pairs of
$\btos$ and $\btod$ three-body decays that are related by U spin. As
pointed out in Refs.~\cite{Gronau:2000zy,Uspinbreak}, the decay rates
and direct CP asymmetries for each U-spin pair satisfy
Eq.~(\ref{Uspinobs}).  Two examples, Eqs.~(\ref{ratio1}) and
(\ref{ratio2}), have already been tested by the asymmetries of
Eqs.~(\ref{asym1}) and (\ref{asym2}), but for the unsymmetrized final
states. One thing we will add is that the prediction of
Eq.~(\ref{Uspinobs}) holds at each point of the Dalitz plot.  Thus,
for the fully-symmetric final states of the U-spin pairs, one should
average over all points.  That is, in the presence of U-spin breaking,
Eq.~(\ref{Uspinobs}) is modified to be
\beq
-\frac{A_s}{A_d} \, \frac{B_s}{B_d} = Y ~,
\eeq
where $Y$ is the U-spin-breaking factor. It is a real number that can
take different values at different points in the Dalitz plot. If $Y >
1$ at some points, and $Y < 1$ at others -- and this will be
determined experimentally -- then averaging over all Dalitz-plot
points will reduce the effect of U-spin breaking, as well as the
statistical error.

\subsection{\boldmath Neglect of $E$/$A$/$PA$}

When $E$/$A$/$PA$ diagrams are neglected, one finds some new
two-amplitude equalities. These provide a good test of the assumption:
\bea
A(B^+ \to K^+ K^+ K^-)_{\rm FS} &=& \sqrt{2} A(\bd \to K^+ K^0 K^-)_{\rm FS} ~, \nn\\
\sqrt{2} A(B^+ \to K^+ K^0 \kbar)_{\rm FS} &=& A(\bd \to K^0 K^0 \kbar)_{\rm FS} ~, \nn\\
{\cal A}(\bs \to \piz\pip\pim)_{\rm FS} & = & 0 ~, \nn\\
A(\bs \to \ok\pi^+\pi^-)_{\rm FS} & = & A(B^+ \to \pi^+K^+K^-)_{\rm FS} ~.
\eea
The \babar\ collaboration has studied the decays $B^+ \to K^+ K^+ K^-$
and $\bd \to K^+ K^0 K^-$, as well as $B^+ \to K^+ K^0 \kbar$ and $\bd
\to K^0 K^0\kbar$ \cite{Expt}. \babar\ data can be used to construct
the fully-symmetric states in these processes to test the first two of
the above relationships.

\section{CONCLUSIONS}

In charmless $B\to PPP$ decays, flavor SU(3) treats the three
final-state particles as identical. As there are six permutations of
these particles, there are six possibilities for the final state: a
totally symmetric state, a totally antisymmetric state, or one of four
mixed states. In this paper, we examine the properties of the
fully-symmetric final state for the case where the final-state
particles are all $\pi$'s or $K$'s.

We begin by writing all $B\to PPP$ decay amplitudes as a function of
the SU(3) reduced matrix elements. There are seven independent
combinations of these matrix elements. On the other hand, there are 16
$\btos$ and 16 $\btod$ decays, for a total of 32. We work out the 25
relations among the amplitudes in the SU(3) limit. Several of these
can be tested experimentally.

We also present all $B\to PPP$ decay amplitudes as a function of
diagrams. By comparing the two expressions for the amplitudes, we are
able to write the matrix elements as a function of diagrams,
demonstrating the equivalence of diagrams and SU(3). One of the
advantage of using diagrams is that it provides dynamical input. In
particular, three of the diagrams -- $E$, $A$ and $PA$ -- all involve
the interaction of the spectator quark, and are expected to be
considerably smaller than the other diagrams. If $E$/$A$/$PA$ are
neglected, there are additional relations among the amplitudes, some
of which can also be tested experimentally.

One relation that provides a good test of the SM is the following. In
the SU(3) limit, one has the equality between two $\btos$ decay
amplitudes: $\sqrt{2} {\cal A}(B^+\to K^+\pi^+\pi^-)_{\rm FS} = {\cal
  A}(B^+\to K^+K^+K^-)_{\rm FS}$. That is, the amplitudes for the
fully-symmetric state of these two decays are predicted to be equal at
each point in the Dalitz plot. Now, LHCb has already measured the
Dalitz plots for these decays. An isobar analysis of the Dalitz plots
allows the fully-symmetric amplitudes to be constructed, so that this
equality, and the SM, can be tested. It is important to average over
all Dalitz-plot points. This reduces the statistical error, and
possibly even the effect of SU(3) breaking. A similar analysis can be
done for the equality between two $\btod$ decay amplitudes: $\sqrt{2}
{\cal A}(B^+\to \pi^+K^+K^-)_{\rm FS} = {\cal A}(B^+\to
\pi^+\pi^+\pi^-)_{\rm FS}$. These tests can be done now; it is hoped
that LHCb will carry out these analyses.

\bigskip
\noindent
{\bf Acknowledgments}: This work was financially supported by the IPP (BB),
by NSERC of Canada (BB, DL), by FQRNT du Qu\'ebec (MI), and by the United
States Department of Energy through Grant No.\ DE FG02 13ER41958 (JLR).


\begin{thebibliography}{99}

\bibitem{3body1} N.~Rey-Le Lorier, M.~Imbeault and D.~London,
  Phys.\ Rev.\ D {\bf 84}, 034040 (2011)
  [arXiv:1011.4972 [hep-ph]].

\bibitem{3body2} M.~Imbeault, N.~Rey-Le Lorier and D.~London,
  Phys.\ Rev.\ D {\bf 84}, 034041 (2011)
  [arXiv:1011.4973 [hep-ph]].

\bibitem{3body3} N.~Rey-Le Lorier and D.~London,
  Phys.\ Rev.\ D {\bf 85}, 016010 (2012)
  [arXiv:1109.0881 [hep-ph]].

\bibitem{BIL} B.~Bhattacharya, M.~Imbeault and D.~London,
  Phys.\ Lett.\ B {\bf 728}, 206 (2014)
  [arXiv:1303.0846 [hep-ph]].

\bibitem{LHCbDS1} R. Aaij {\it et al.}  [LHCb Collaboration],
  Phys.\ Rev.\ Lett.\  {\bf 111}, 101801 (2013)
  [arXiv:1306.1246 [hep-ex]].

\bibitem{Aaij:2013bla}
  R.~Aaij {\it et al.}  [LHCb Collaboration],
  Phys.\ Rev.\ Lett.\  {\bf 112}, 011801 (2014)
  [arXiv:1310.4740 [hep-ex]].

\bibitem{Bhattacharya:2013cvn}
  B.~Bhattacharya, M.~Gronau and J.~L.~Rosner,
  Phys.\ Lett.\ B {\bf 726}, 337 (2013)
  [arXiv:1306.2625 [hep-ph]].

\bibitem{He}
D.~Xu, G.~-N.~Li and X.~-G.~He,
arXiv:1307.7186 [hep-ph], arXiv:1311.3714 [hep-ph].

\bibitem{MG}
M.~Gronau,
  Phys.\ Lett.\ B {\bf 727}, 136 (2013)
  [arXiv:1308.3448 [hep-ph]].

  \bibitem{Cheng}
H.~-Y.~Cheng and C.~-K.~Chua,
  Phys.\ Rev.\ D {\bf 88}, 114014 (2013)
  [arXiv:1308.5139 [hep-ph]].

\bibitem{GHLR1}  M.~Gronau, O.~F.~Hern\'andez, D.~London and J.~L.~Rosner,
  Phys.\ Rev.\ D {\bf 50}, 4529 (1994)
  [hep-ph/9404283].

\bibitem{deSwart:1963}
  J.~J.~de Swart,
  Rev.\ Mod.\ Phys.\  {\bf 35}, 916 (1963)
  [Erratum-ibid.\  {\bf 37}, 326 (1965)].

\bibitem{Kaeding} T.~A.~Kaeding,
  nucl-th/9502037.

\bibitem{McNamee:1964} P.~S.~J.~McNamee and F.~Chilton,
  Rev.\ Mod.\ Phys.\  {\bf 36}, 1005 (1964).

\bibitem{Zeppenfeld} D.~Zeppenfeld,
  Z.\ Phys.\ C {\bf 8}, 77 (1981).

\bibitem{GRBKKK} M.~Gronau and J.~L.~Rosner,
  Phys.\ Lett.\ B {\bf 564}, 90 (2003)
  [hep-ph/0304178].

  \bibitem{Gronau:2000zy}
M.~Gronau,
  Phys.\ Lett.\ B {\bf 492}, 297 (2000)
  [hep-ph/0008292].
.
\bibitem{Uspinbreak} M.~Imbeault and D.~London,
  Phys.\ Rev.\ D {\bf 84}, 056002 (2011)
  [arXiv:1106.2511 [hep-ph]].

\bibitem{GHLR2}  M.~Gronau, O.~F.~Hern\'andez, D.~London and J.~L.~Rosner,
  Phys.\ Rev.\ D {\bf 52}, 6374 (1995)
  [hep-ph/9504327].

\bibitem{NR} M.~Neubert and J.~L.~Rosner,
  Phys.\ Lett.\  B {\bf 441}, 403 (1998)
  [arXiv:hep-ph/9808493],
  Phys.\ Lett.\  B {\bf 441}, 403 (1998)
  [arXiv:hep-ph/9808493].

\bibitem{GPY} M.~Gronau, D.~Pirjol and T.~M.~Yan,
  Phys.\ Rev.\  D {\bf 60}, 034021 (1999)
  [Erratum-ibid.\  D {\bf 69}, 119901 (2004)]
  [arXiv:hep-ph/9810482].

  %
  \bibitem{PDG} J. Beringert {\it et al.} (Particle Data Group), Phys. Rev. D {\bf 86} (2012) 01000,
  partially updated for the 2014 edition, http://pdg8.lbl.gov/rpp2013v2/pdgLive/Viewer.action.

\bibitem{Expt}
J.~P.~Lees {\it et al.}  [BABAR Collaboration],
  Phys.\ Rev.\ D {\bf 85}, 112010 (2012)
  [arXiv:1201.5897 [hep-ex]];
  J.~P.~Lees {\it et al.}  [BABAR Collaboration],
  Phys.\ Rev.\ D {\bf 85}, 054023 (2012)
  [arXiv:1111.3636 [hep-ex]].

\end{thebibliography}
\end{document}